\documentclass{JINST}

\usepackage[latin1]{inputenc}
\usepackage{subfigure}
\usepackage{fontenc}
\usepackage{amsmath}
\usepackage{url}

\title{W-band prototype of platelet feed-horn array for CMB polarisation measurements}

\author{
F.~Del Torto$^a$\thanks{Corresponding author.}, 
M.~Bersanelli$^a$, 
F.~Cavaliere$^a$, 
A.~De Rosa$^b$, 
O.~D'Arcangelo$^c$,
C.~Franceschet$^a$, 
M.~Gervasi$^d$, 
A.~Mennella$^a$, 
E.~Pagana$^e$, 
A.~Simonetto$^c$, 
A.~Tartari$^d$, 
F.~Villa$^b$, 
M.~Zannoni$^d$\\\llap{$^a$}Dipartimento di Fisica, Universit\`a degli Studi di Milano,\\
via Celoria 16, 20133 Milano, Italy,\\
\llap{$^b$}INAF-IASF Bologna,\\ 
Via Gobetti 101, 40129 Bologna, Italy,\\
\llap{$^c$}IFP-CNR,\\ Via Cozzi 53, Milano, Italy\\ 
\llap{$^d$}Dipartimento di Fisica, Universit\`a degli Studi di Milano Bicocca,\\
Piazza della Scienza 3, 20126 Milano, Italy,\\
\llap{$^e$}Independent consultant\\ 

E-mail: \email{francesco.deltorto@mi.infn.it}
}

\abstract{
We present the design and performance of a 2x2 prototype array
of corrugated feed-horns in W-band. The module is fabricated using a so-called ``platelet'' technique by milling Aluminum plates. This technique is suitable for low-cost and scalable high performance applications. Room temperature Return Loss measurements show a low (<-30 dB) reflection over a 30\% bandwidth with a maximum matching of -42 dB at 100 GHz for all four antennas. Beam pattern measurements indicate good repeatability and a low (-25 dB) sidelobe and crosspolarisation levels. This work is particularly relevant for future Cosmic Microwave Background polarisation measurements, which require large microwave cryogenic detector arrays coupled to high performance corrugated feed horns.
}


\begin{document}

%%%%%%%%%%%%%%%%%%%%%%%%%%%%%%%%%%%% SECTION 1 %%%%%%%%%%%%%%%%%%%%%%%%%%%%%%%%%%%%

\section{Introduction}
Forty-five years after its discovery, the \textbf{C}osmic \textbf{M}icrowave \textbf{B}ackground (CMB) radiation is today a very powerful tool of cosmological investigation. 

The CMB originates from the decoupling of matter and radiation, occurred when the temperature of the expanding universe fell below 3000\,K. At this temperature the plasma could combine into neutral hydrogen, thus reducing the Thomson scattering cross section and letting the radiation propagate freely.
Because of cosmic expansion the CMB can be observed today as a blackbody radiation at the temperature T=2.725 $\pm$ 0.002 K \cite{Mather} peaked in the microwave region of the electromagnetic spectrum.

The CMB temperature is anisotropic at the level of $\Delta T/T\approx10^{-5}$ on scales from 180 degrees to about $5^{\prime}$. The angular temperature anisotropies depend on the initial conditions of the Universe and on the physical processes that occurred from the Big Bang to the matter-radiation decoupling and later via Sunyaev-Zel'dovich effect. Accurate measurements of these anisotropies, therefore, provide a wealth of information on the universe history and composition (baryons, dark matter, dark energy).
CMB temperature anisotropies have been measured by many ground and balloon experiments as well as by the NASA satellites COBE\footnote{\url{http://aether.lbl.gov/www/projects/cobe/}} and WMAP\footnote{\url{http://map.gsfc.nasa.gov/}} (see  \cite{Anisotropies} for an extensive review until WMAP). The third generation ESA satellite Planck\footnote{\url{http://planck.esa.int}}, launched in May 2009, will provide in few years foreground-limited CMB temperature anisotropy measurements and open the path to accurate measurements of CMB polarisation anisotropies, the future of CMB experiments.

Polarisation anisotropies, generated by Thomson scattering, can be represented on the celestial sphere by two modes, named E-mode and B-mode, according to their symmetry properties. The detection of B-modes would provide an indirect signature of cosmological gravitational waves, that are predicted by the Hot Big Bang model \cite{Pol_Primer}.

Many experiments are currently planned to accurately measure the CMB polarisation. First detection of E-modes and its correlation with the temperature anisotropy was published in 2003 by the DASI collaboration \cite{DASI_and_others}, while the most up-to-date measurements come from WMAP \cite{WMAP_pol}.
Because only 10\% of the CMB is polarised, polarisation anisotropies are smaller by a factor 10 compared to temperature anisotropies.
E-modes, in particular, are at the $\mu$K level, while for B-modes the signal is expected to be much weaker and not even well theoretically constrained, although it is not exptected to be larger than some fraction of $\mu$K \cite{Pol_Primer}.

Detection of such small signals call for ultra-high sensitivity instruments constituted of large detector arrays of hundreds and even thousands of channels. A key element is represented by the availability of high performance antennas which represent the interface between the sky (or telescope) and the receiver. Precision polarisation measurements require optical components with highly symmetrical optical response and losses not greater than $\sim$1\%. Corrugated feed-horns match these requirements and are commonly used in most CMB experiments [cfr. WMAP and Planck]. Typical manufacturing techniques include direct machining, up $\sim$40 GHz, and electroforming, at higher frequencies. These methods can be limited by time and cost when applied to the production of large detector arrays of hundreds of feed-horns, as requested by future CMB polarisation experiments, so cheaper and quicker techniques are required.

In this work we investigate the validity of a technique, named ``platelet'' \cite{Kangas}, \cite{Haas}, \cite{Wollack}, for the mass production of corrugated feed-horn arrays. In particular we focus on three key factors: (i) mechanical feasibility with reduced costs and time, (ii) reliability in standard and cryogenic conditions and (iii) repeatability of the electromagnetic horn performance. Recently the QUIET\footnote{\url{http://quiet.uchicago.edu}} experiment started CMB polarisation measurememnts using, for the first time, feed-horn array realized with the platelet technique \cite{QUIET2010}.

The platelet technique consists in the overlap of properly machined metal sheets so that the feed-horn is then realized in layers. This method is relatively cheap, with a cost saving of about 90\% with respect to the electroforming and allows quick manufacturing of feed-horn arrays by drilling several holes in each plate.
A complete corrugation, with a tooth and a groove, can be obtained in each plate.
A complex focal surface can be obtained with a series of modules properly tilted to reproduce the required geometry.
Particular care is required in the plate assembly alignment system and for mass/volume control in order to minimize the weight. 

The advantages of this technique go beyond its application to CMB measurements and can be exploited in a wide range of mm-wave and sub-mm astronomical instruments. In particular, platelet techniques may prove appropriate for the new generation of large interferometric arrays, such as ALMA, which require high performance antennas (including polarisation properties) over a large number of channels.

%%%%%%%%%%%%%%%%%%%%%%%%%%%%%%%%%%%% SECTION 2 %%%%%%%%%%%%%%%%%%%%%%%%%%%%%%%%%%%%

\section{The Prototype}
To validate the ``platelet'' technique we have realized a prototype of 2x2 array of corrugated feed-horn in W band (75-110 GHz), named WCAM01.

\subsection{Electromagnetic design}
The antenna electromagnetic design started form an already existing one \cite{Villa97}. The design was scaled to W band and the corrugation step was fixed to 1 mm in order to use standard aluminum plates. The circular waveguide radius at the throat region was adapted by means of three steps from 1.87 mm to 1.49 mm in order to match the circular-rectangular transition. In addition a 4 mm long straight section has been added to guarantee the propagation of the TE11 mode at the throat. Because the main objective of this development was to test the manufacturing technique, we did not perform any further optimizations of the electromagnetic design.

\begin{figure}[!h]
\centering
	\includegraphics[width=0.8\textwidth]{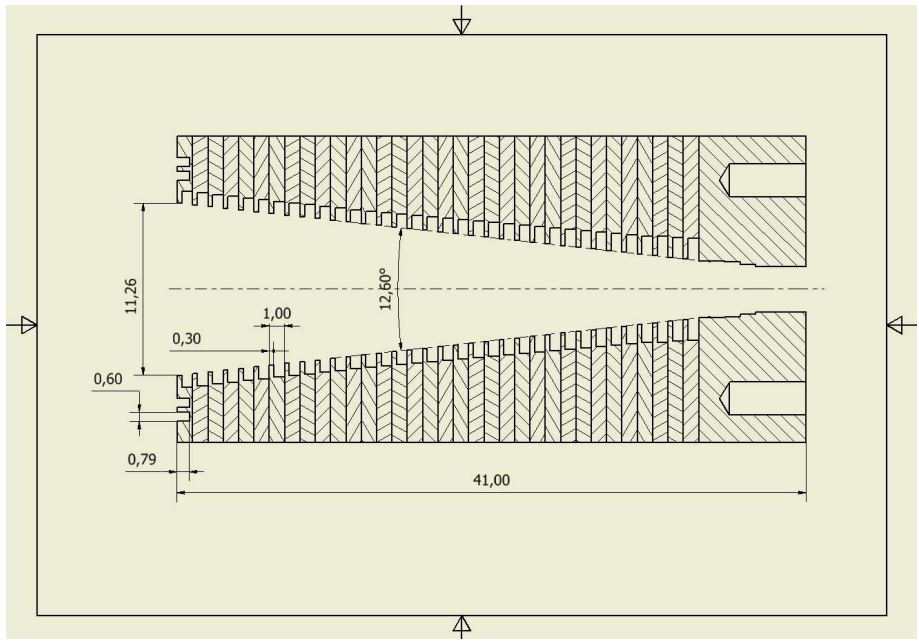}
	 \caption{Section of a corrugated feed-horn of the WCAM01 prototype.}
	\label{fig:sezione}
\end{figure}

The horn aperture is $D = 11.26$~mm, giving a FWHM ($\sim\lambda/D\,$) of $\sim 18$~degrees at 100~GHz. Its length is 41~mm with a flare angle of 12.6~degrees. Each corrugation tooth and groove is 0.3~mm and 0.7~mm wide, respectively. The corrugation depth is $\sim\lambda/4$ at the aperture and linearly increases to $\sim\lambda/2$ in the last seven corrugations. A section representing the horn profile is shown in Figure \ref{fig:sezione}.

The beam pattern simulations were obtained with the SRSR-D$^\copyright$ France Telecom software. The software is designed to simulate the electromagnetic performance of structures with cylindrical symmetry and consisting of perfect electrical conductor parts and homogeneous dielectric domains, therefore neglecting the metal resistive losses (i.e. the insertion loss). These restrictions result in some limitations but allow very fast and reliable simulations. 
The radiation patterns have been calculated at 90\,GHz, 92.5\,GHz, from 95\,GHz to 105\,GHz with a step of 0.25\,GHz, then at 107.5\,GHz and 110\,GHz. Return loss was simulated with CST Microwave Studio$^\copyright$ software.

\subsection{Mechanical design}

The mechanical design was conceived with the driving idea to obtain scalable modules suitable to realise large feed-horn  arrays. The prototype was manufactured (Figure \ref{fig:realizzazione}) with a CNC machine tool at the Physics Department workshop of the University of Milan.

The array has a square section of 70$\times$70\,mm, it is 44\,mm high and composed by:

\begin{itemize}
\item {one 7-mm thick base plate,}
\item {thirty-three 1-mm thick plates,}
\item {one 4-mm thick head plate.}
\end{itemize}
The plates are aligned by means of 8 rectified pins, alternated in pairs, and assembled together with 9 rods. In this prototype the rod number and displacements have not been optimized for high packing efficiency (horn aperture area/total area $=0.08$) and lightweight. This will be the objective of future studies aiming at designing more compact and light modules.
The base plate is 7 mm thick in order to host the three following elements: (i) the three steps required to reach the optimal exit diameter of 2.98 mm, (ii) the anchorage holes (on the external side) for screws and pins for the standard UG387 flanges, and (iii) the holes for the first alignment pins, that must be deep enough to ensure that they are perpendicular to the plate.
The 4 mm thickness of the head plate is required to hide the rod nuts. 

The feed-horn positioning and the interfaces were not designed for a particular type of detector but just compatible with standard WR-10 waveguide flanges. In principle the array could feed both bolometric detectors (with an appropriate thermal brake) and MMIC amplifier based detectors.

\subsection{Materials}
The chosen materials have been two aluminum alloys: Al6082 Al-Mg-Si-Mn (Anticorodal) for the plates, and Al70750 Al-Zn-Mg-Cu (Ergal) for the rods. The pins and the nuts are made of stainless steel.

The main characteristics of these aluminum alloys are their low specific weight (2.69 g/cm$^{3}$ for the Al 6082, 2.81 g/cm$^{3}$ for the Al 7075), their excellent workability and corrosion resistance, their ability to be anodized and their good strength. The main difference between Al6082 and Al7075 is the tensile strength that is respectively 310 N/mm$^{2}$ and 560 N/mm$^{2}$. Moreover their thermal expansion coefficients are quite similar: 23.4 for the Al6082 and 23.5 for the Al7075, thus allowing to apply a great rod traction to properly bind the plates, and to stabilize the prototype with respect to thermal contractions.

\subsection{Manufacturing}
Plates 1 through 33 were cut from 1~mm thick aluminum sheets and worked with the CNC machine tool ``Gualdoni GV-94''. The base-plate and head-plate were milled from a 10 mm thick aluminum slab and worked with the same machine. The rods were obtained by lathing 8~mm diameter Ergal bars, threaded at their ends.

\begin{figure}[h!]
 \centering
	\includegraphics[angle=0,width=0.45\textwidth,]{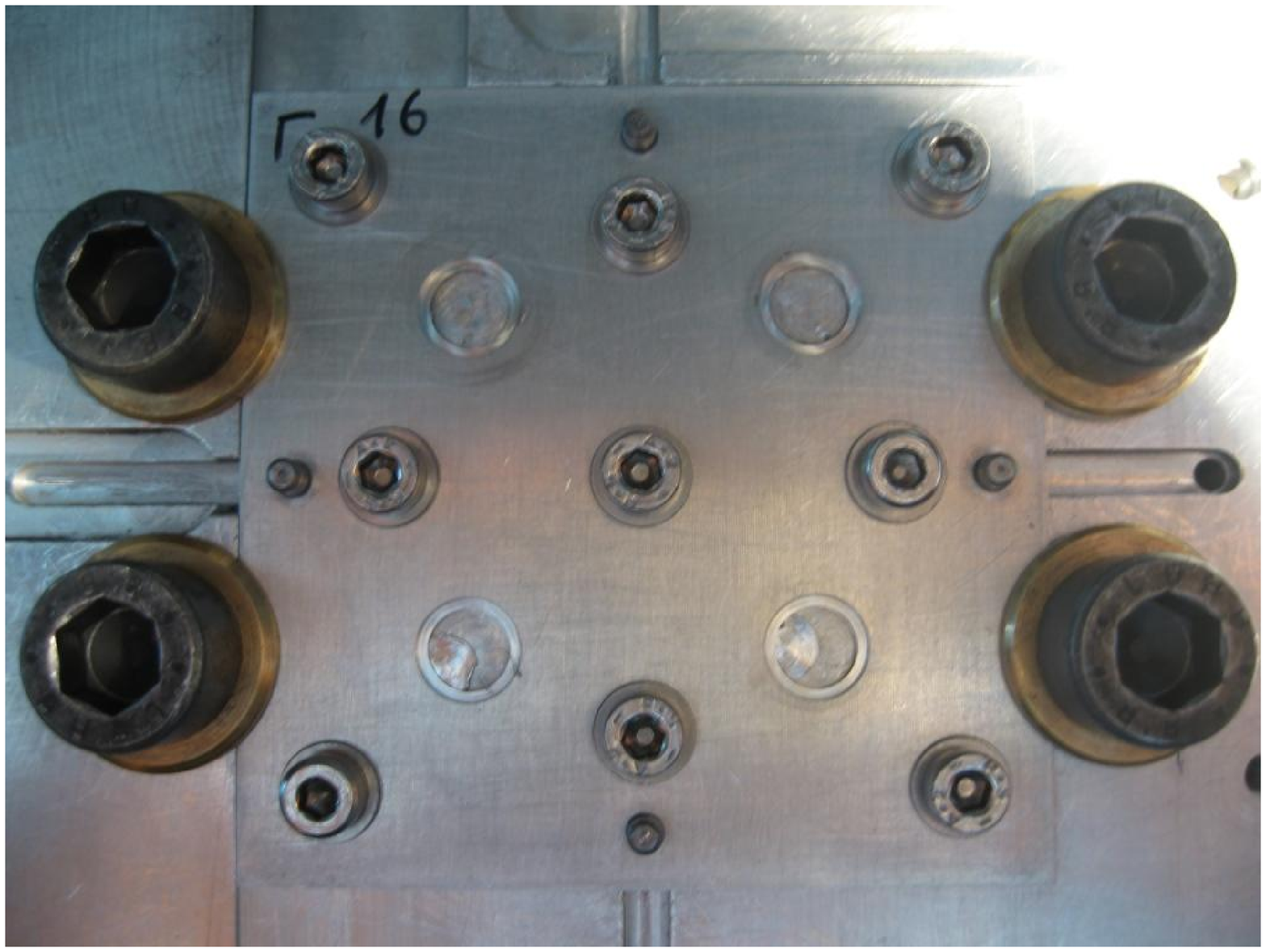}
	\includegraphics[angle=0,width=0.45\textwidth]{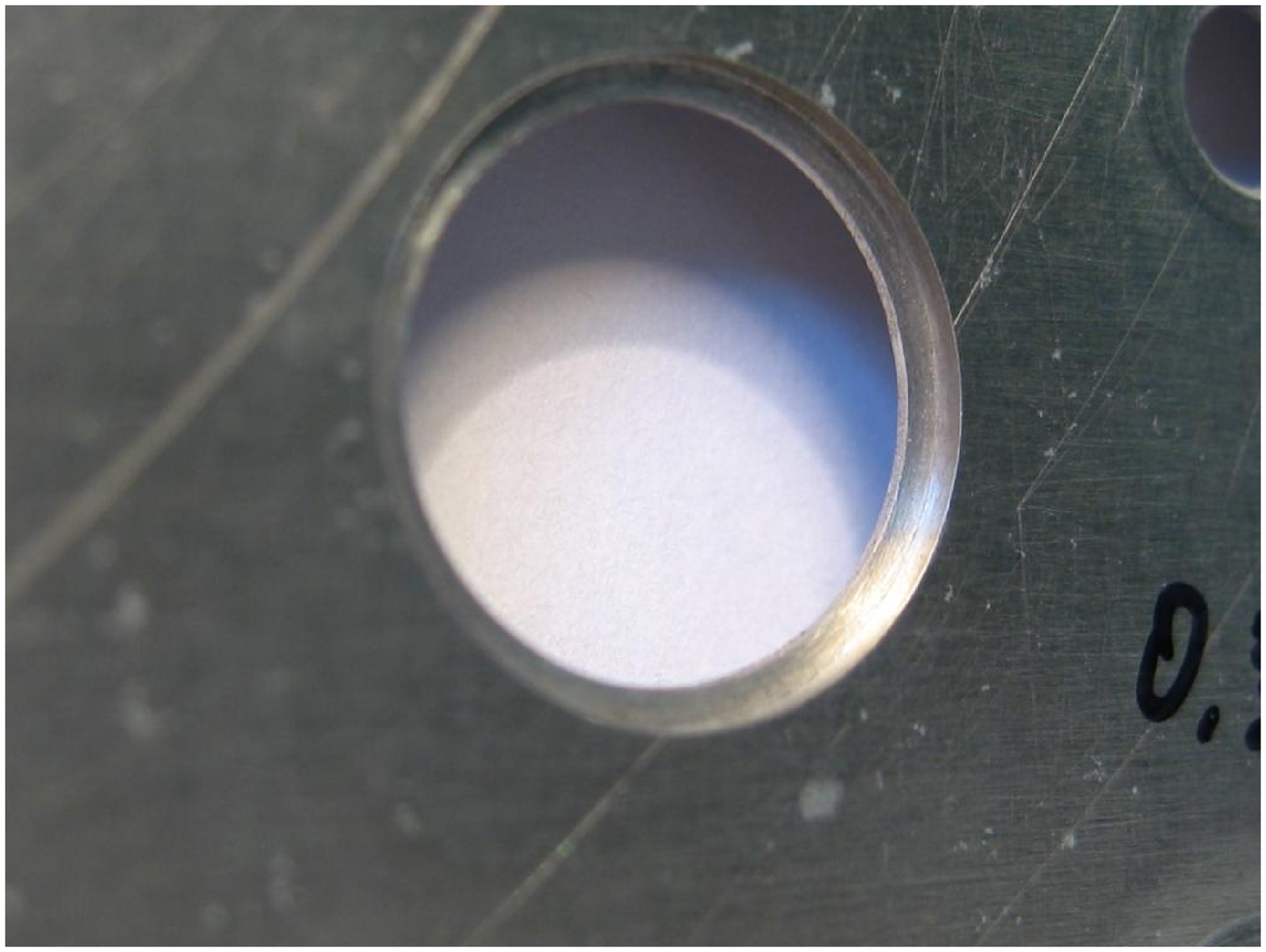}
	\includegraphics[angle=0,width=0.45\textwidth]{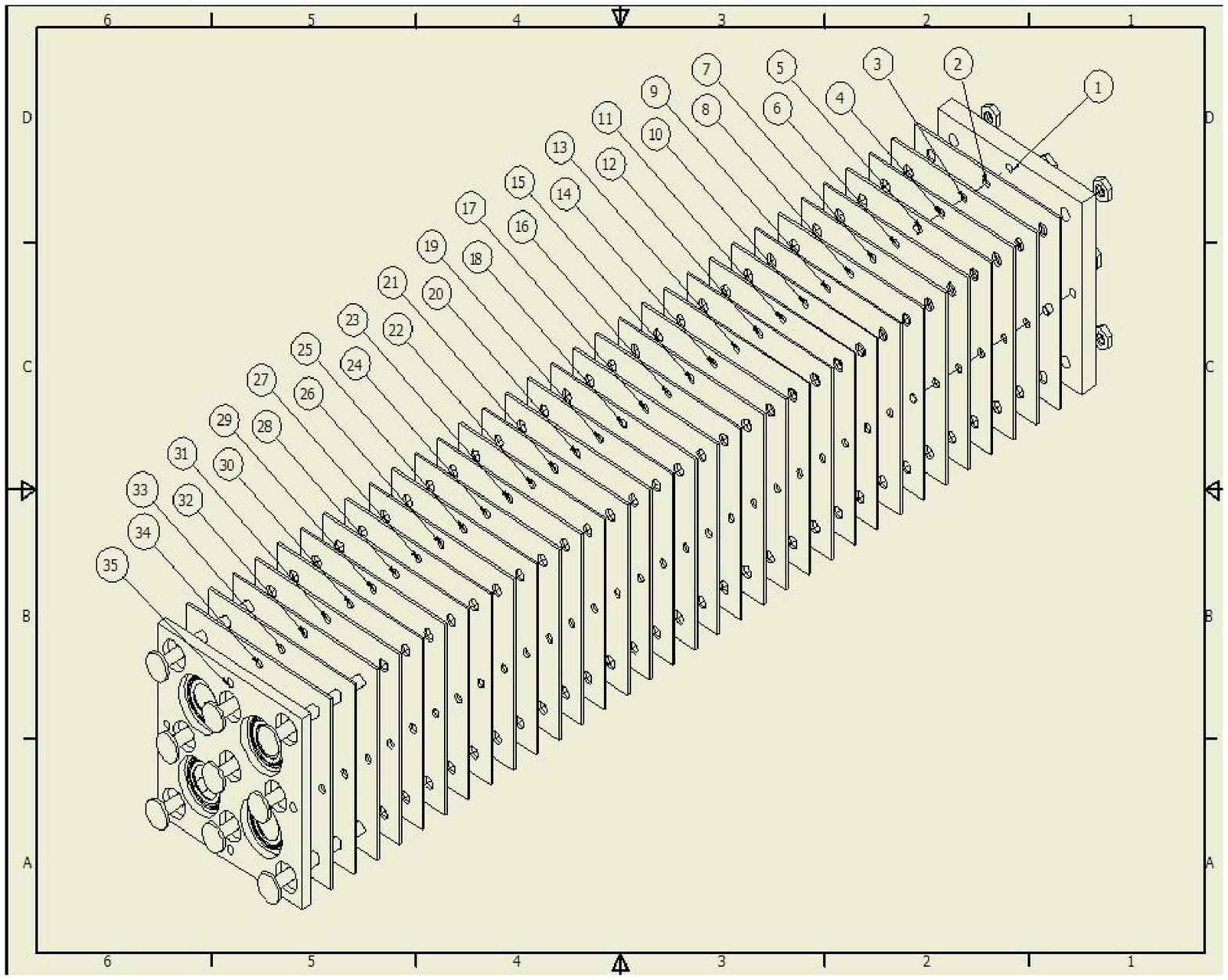}
	\includegraphics[angle=0,width=0.45\textwidth]{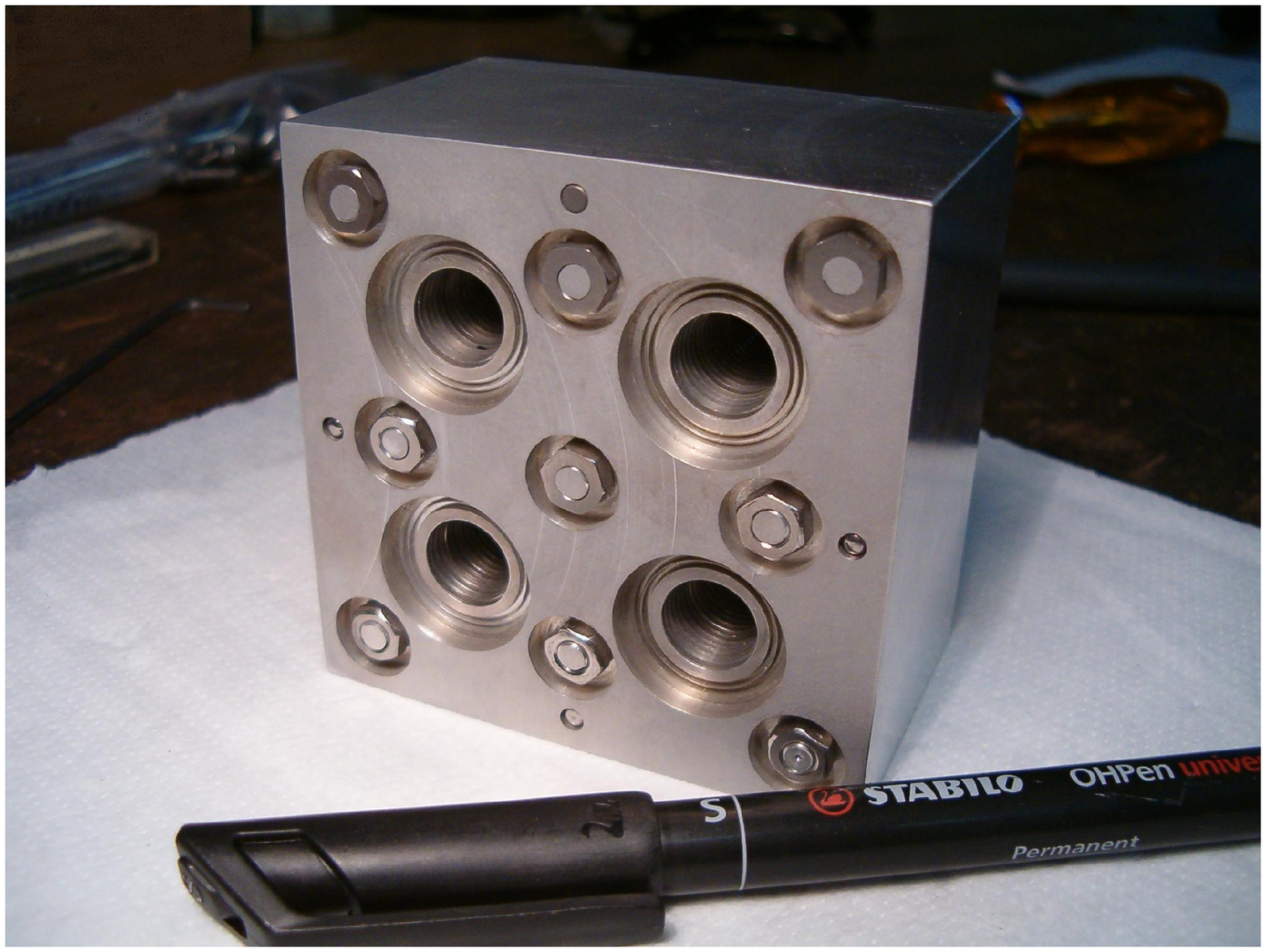}
	 \caption{Top Left: top view of a single plate screwed inside the CNC. There are nine screws: four (black) on the sides and five (gray) located in previously drilled rod-holes. After fixing the plate the four antenna holes were milled. Top Right: corrugation details. Bottom Left: prototype assembling scheme. Bottom Right: The completed prototype.}
	\label{fig:realizzazione}
\end{figure}

The corrugations were obtained by milling circular hollows concentric to the holes (Figure \ref{fig:realizzazione}), with the difference in diameter corresponding to the corrugation depth, and the cave depth (0.7~mm) corresponding to the the corrugation groove thickness. 
This process requested particular care to ensure the orthogonality between the plates and the mechanical tool. This condition was guaranteed by a custom-designed plate support. The maximum error obtained on the teeth corrugations thickness was of $\pm$30~$\mu$m.

Finally, the plates were manually assembled and aligned with the steel pins, and tightened with the Ergal bolts. The assembly by means of rods is very quick, simple and does not require any facilities and/or machinery. Moreover the array can be disassembled and modified in order to possibly change its properties or replace inaccurate and/or damaged parts. Compared to bonding, our technique has the drawback that a smaller number of horns can be packed in a single module. On the other hand its simplicity, the flexibility that comes from the possibility to disassemble the module and the proven performance in cryogenic conditions (see tests discussed in Section \ref{sec_testing}) led us to choose this approach for our prototype.

Total weight is 533 grams, the rods and pins representing the 2.6\% and 1.8\%, respectively.

%%%%%%%%%%%%%%%%%%%%%%%%%%%%%%%%%%% SECTION 3 %%%%%%%%%%%%%%%%%%%%%%%%%%%%%%%%%%%%

\section{Testing}
\label{sec_testing}
Electromagnetic tests were performed to validate the prototype and the manufacturing technique by comparing measurements with electromagnetic simulations. In particular we measured the angular response (beam patterns), insertion loss (IL) and return loss (RL),  which are related to the scattering parameters as follows:
\begin{equation}
	IL=10\cdot \log_{10}\frac{|S_{21}|^{2}}{1-|S_{11}|^{2}}; \hspace{1.5cm}	
	RL=-10\cdot \log_{10}|S_{11}|^{2}.
\end{equation}

\subsection{Beam Patterns}
Beam patterns were measured at 100~GHz in the anechoic chamber of the ``Istituto di Fisica del Plasma'' (IFP-CNR) in Milan with an ABmillimetre 8-350-4 VNA. They were obtained by connecting one horn at a time to one VNA port using a tapered rectangular-to-circular transition. The signal, launched by the horn under test, was collected by a standard gain horn connected to the second VNA port.
The E and H orthogonal planes beam patterns were measured for all four horns, while the 45 degrees plane and the relative cross-polarisation were measured for one horn. 

\begin{figure}[h!]
\centering
	\includegraphics[angle=0,width=0.85\textwidth]{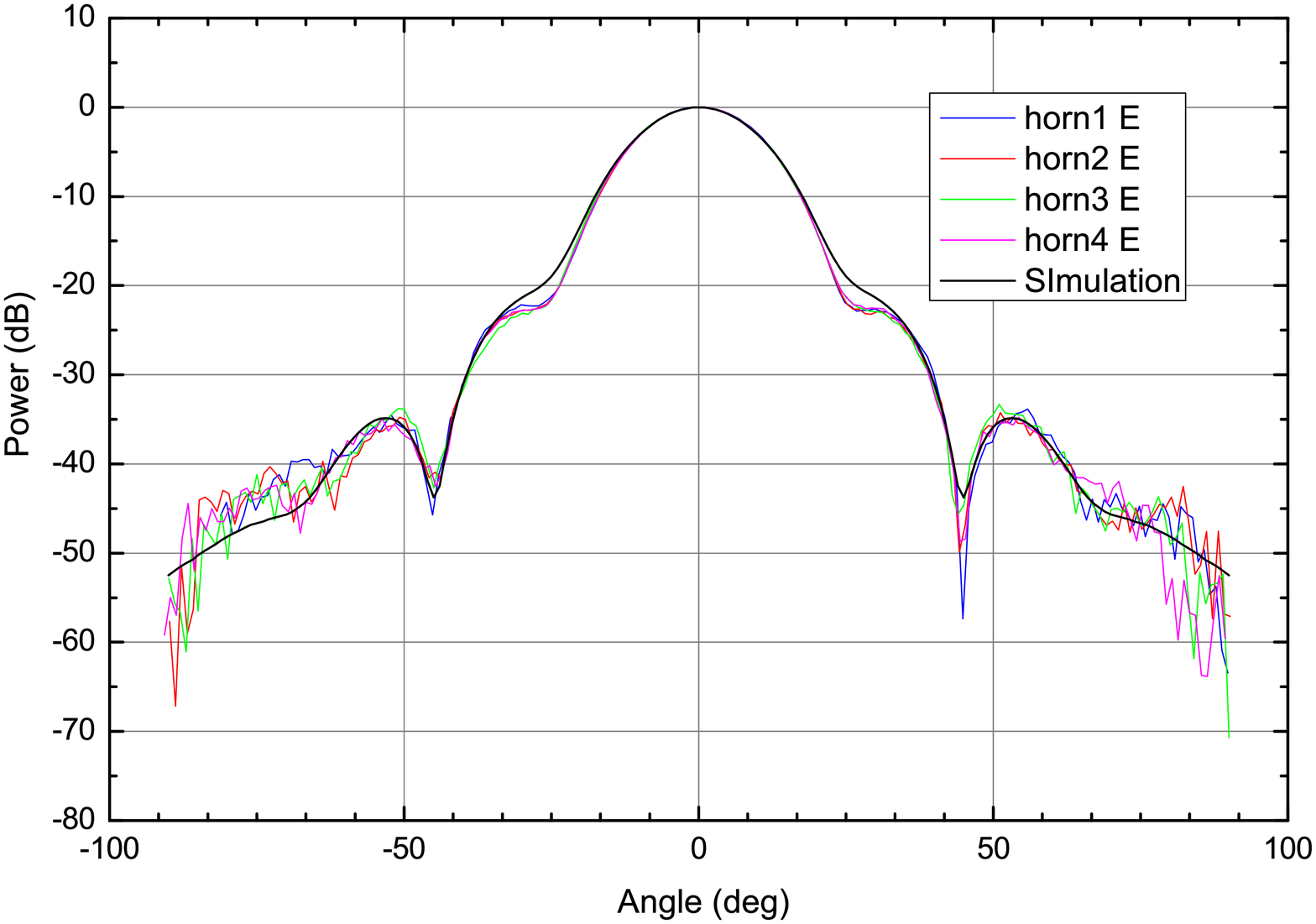}
	 \caption{Comparison of measured and simulated E-plane beam patterns}
	\label{fig:piani_E}
\end{figure}

\begin{figure}[h!]
\centering
	\includegraphics[angle=0,width=0.85\textwidth]{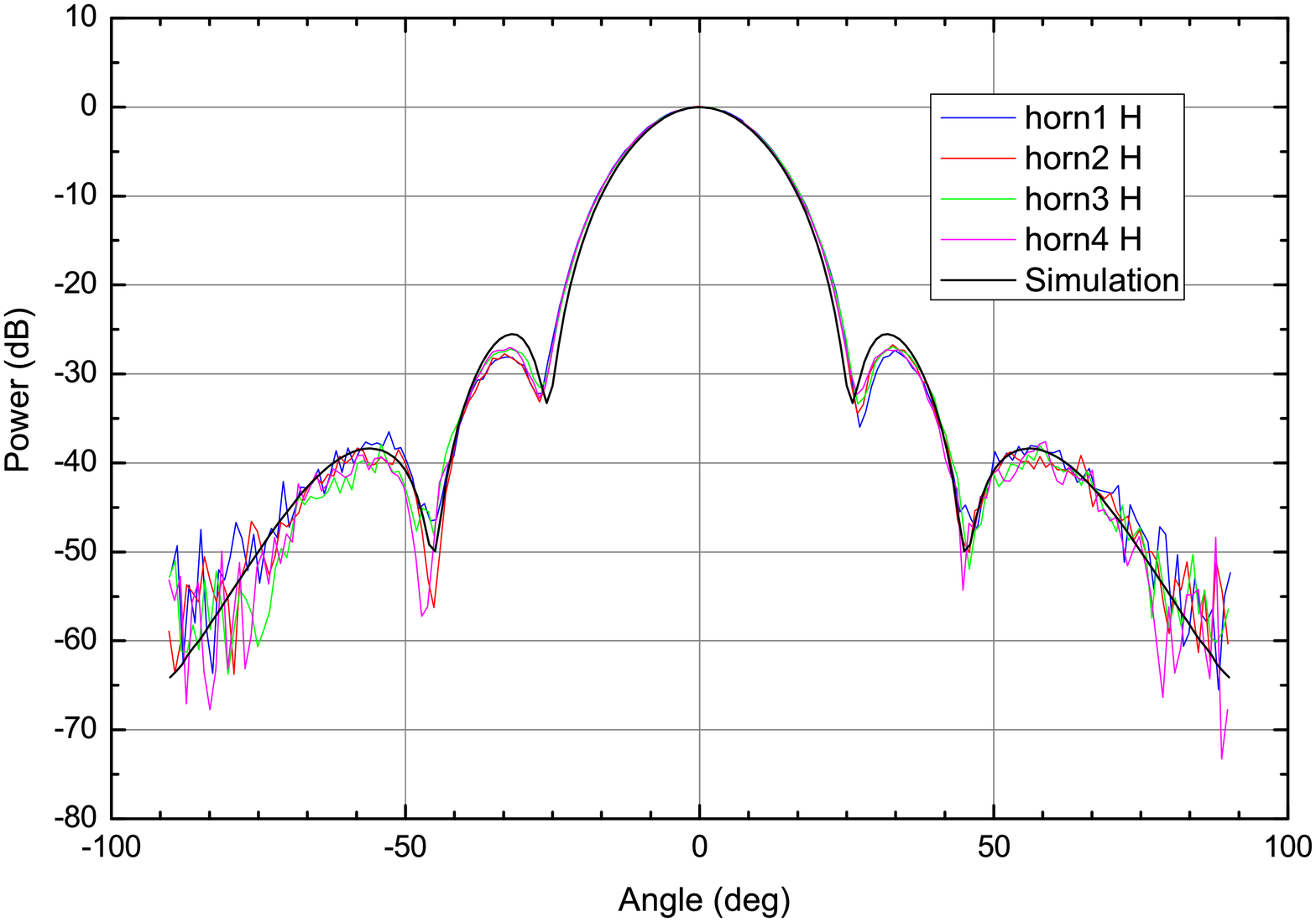}
	 \caption{Comparison of measured and simulated H-plane beam patterns}
	\label{fig:piani_H}
\end{figure}

The measured E-plane and H-plane patterns for all feed horns, compared with the simulations (black line) described in \cite{sim_Villa}, are shown in Figures \ref{fig:piani_E} and \ref{fig:piani_H}, respectively.
 
The four patterns are very similar, both in E and H planes, and in good agreement with simulations. At large angles ($|\theta| > 65^\circ$) the horn response is below -40 dB and the signal to noise ratio in the measurements degrades. The $\sim$ 2dB deviation (at -20 dB level) between measurements and simulations, around the first side lobes, could be due to mechanical manufacturing tolerances, that are of the order of $\sim$ 0.03 mm. A comparison of these results with those of an electroformed equivalent horn could clarify the source of these small deviations.

\begin{figure}[!ht]
\centering
	\includegraphics[angle=0,width=0.85\textwidth]{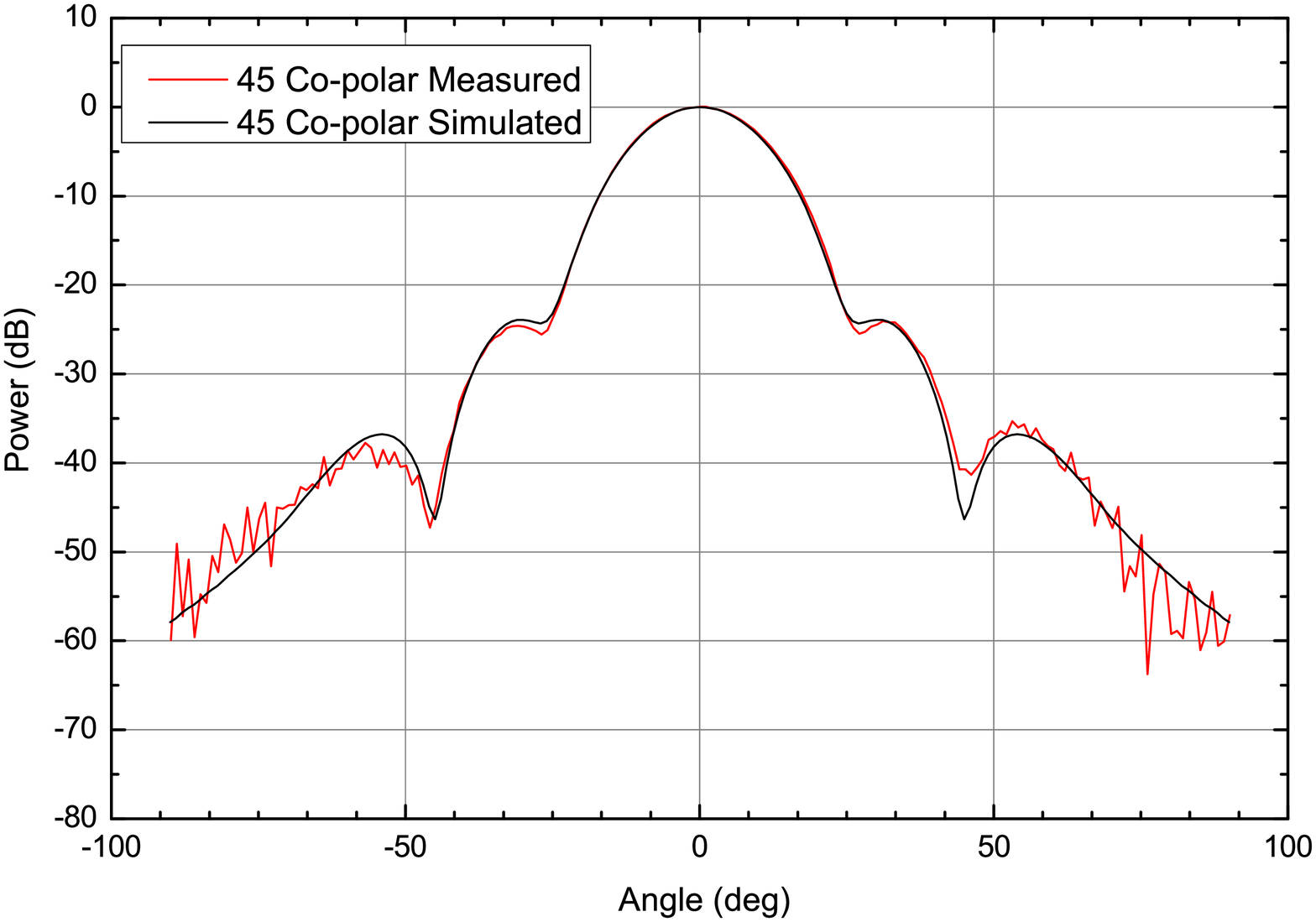}
	 \caption{Comparison of the measured and simulated (black line) patterns of the 45$^\circ$ plane of one horn.}
	\label{fig:piano_45}
\end{figure}

In Figure \ref{fig:piano_45} a comparison between the measured and simulated (black line) patterns of the 45 degrees plane of one horn is shown. The results are similar to the E and H planes cases.

\begin{figure}[!ht]
\centering
	\includegraphics[angle=0,width=0.85\textwidth]{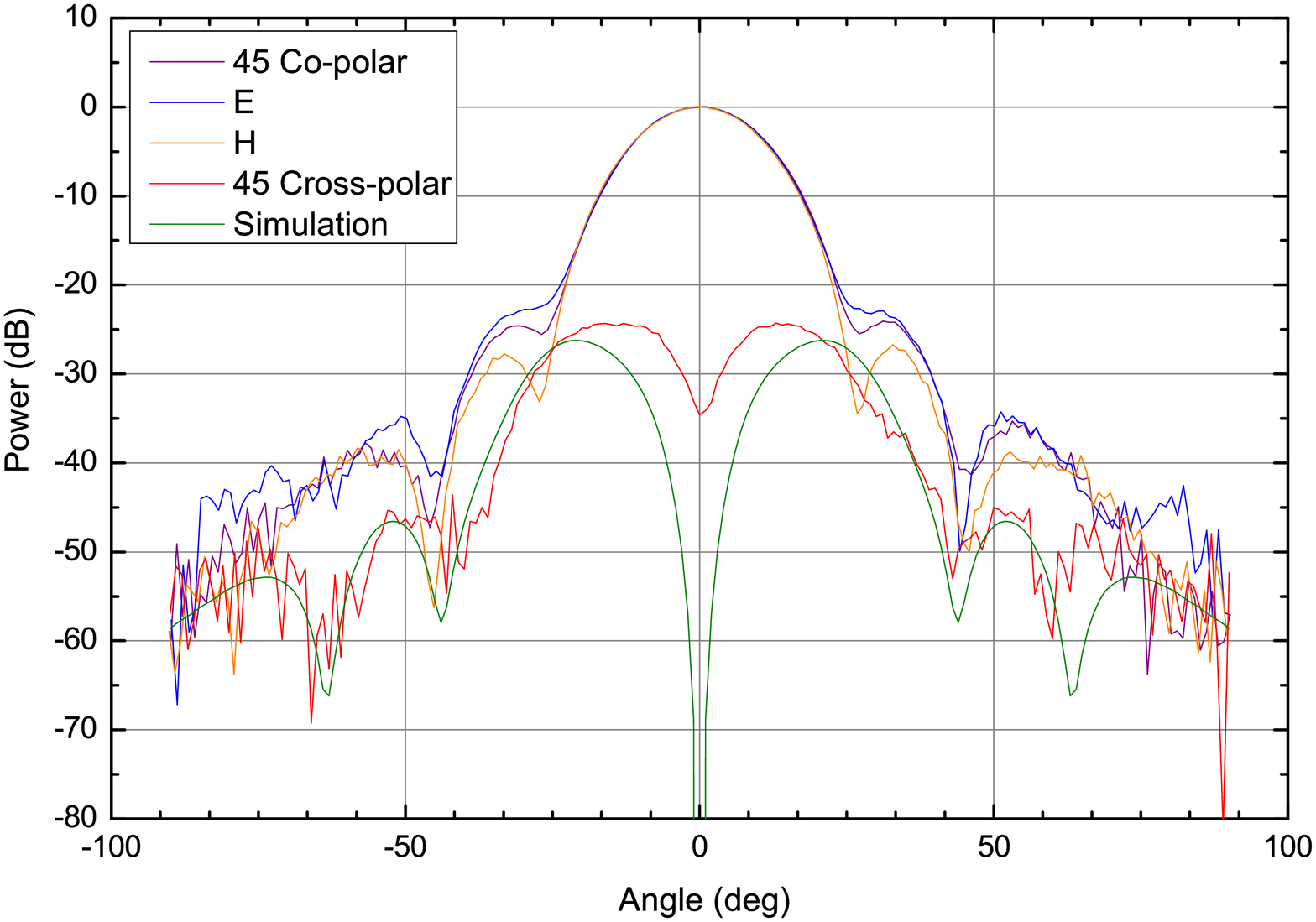}
	 \caption{The planes E, H, 45$^\circ$ co-polar and 45$^\circ$ cross-polar for one the horn. The green line represents the simulation (from \cite{Villa97}) of the 45$^\circ$ cross-polar plane.}
	\label{fig:horn2full}
\end{figure}

Figure \ref{fig:horn2full} shows the comparison between the measured (red line) and simulated (green line) cross-polarisation patterns relative to the 45 degrees plane of one horn. Also in this case deviations could be due to mechanical manufacturing tolerances.

\subsection{Return Loss}
\begin{figure}[!h]
\centering
	\includegraphics[angle=0,width=0.85\textwidth]{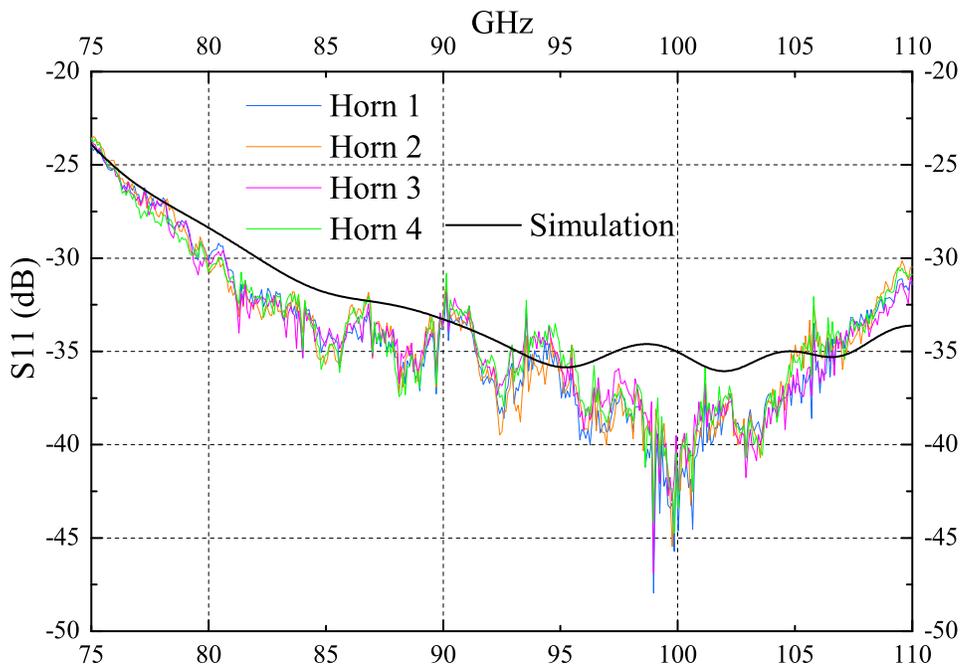}
	 \caption{Return loss measurements compared with the simulation.}
	\label{fig:RL_TRL3}
\end{figure}

The RL measurements were obtained at the Physics Department of the University of Milano-Bicocca with an Agilent 8510c VNA, connecting one horn at a time to one VNA port by means of a tapered rectangular-to-circular transition. The VNA was calibrated with a one-port calibration at the interface between the horn and the transition. The measured return loss is below -30 dB on a 30\% bandwidth, with a maximum matching of -42 dB at 100 GHz, resulting within the simulated requirement (see Figure \ref{fig:RL_TRL3}). Moreover the measurements shows high repeatability for all horns. We recall that the simulation does not take into account resistive losses (see Sec. 3.3), which is compatible with the deviations at the -40 dB level. Moreover the measurements show high repeatability among all horns.

\begin{figure}[!ht]
\centering
	\includegraphics[angle=0,width=0.85\textwidth]{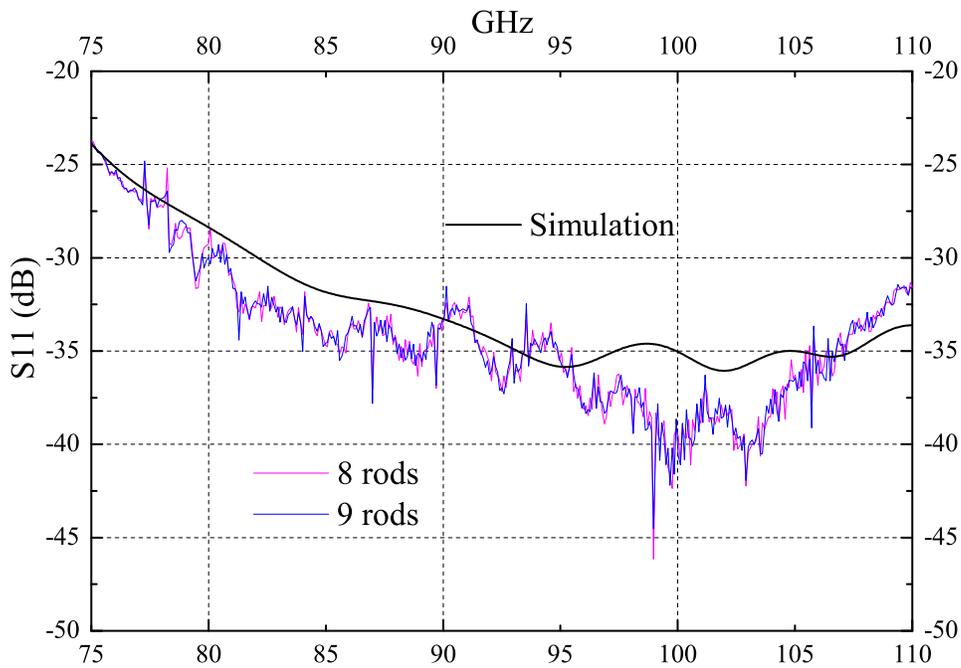}
	 \caption{Comparison of the RL with (blue curve) and without (pink curve) the central rod, the curves are very similar. This result allows, in principle, to substitute the central rod with another horn.}
	\label{fig:8_rods}
\end{figure}

\begin{figure}[!ht]
\centering
	\includegraphics[angle=0,width=0.85\textwidth]{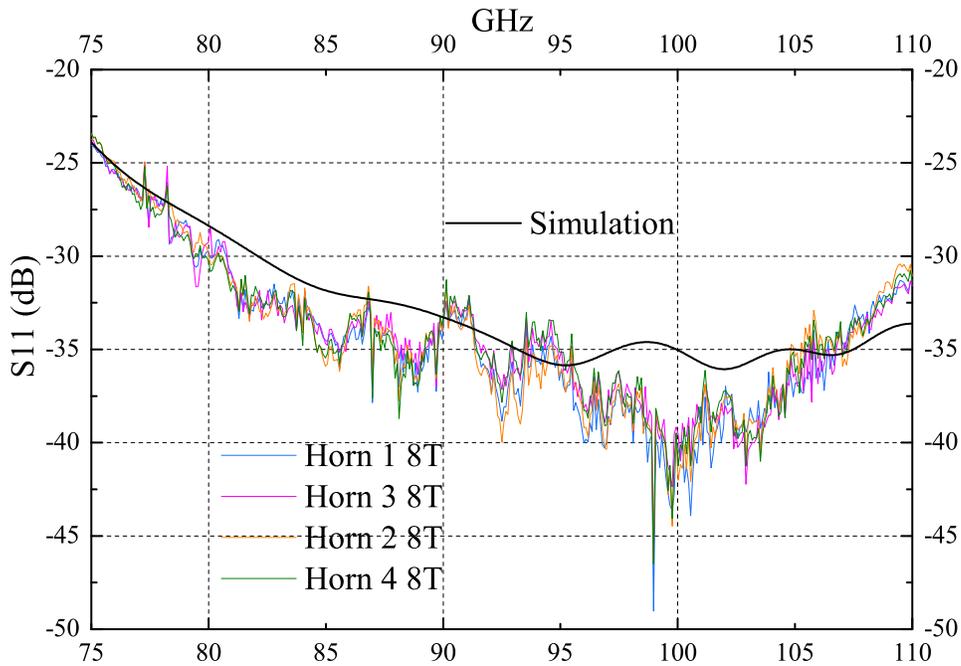}
	 \caption{Return loss measurements of the 4 horns without the central rod. The curves are very similar to the curves obtained with all the nine rods and they show high repeatability for all the four horns.}
	\label{fig:S11_no-tirante}
\end{figure}

Another return loss measurement, performed after removing the central rod, yielded very similar results, (see Figures \ref{fig:8_rods} and \ref{fig:S11_no-tirante}), indicating that it could be possible, in principle, to use this space to build a  module with an increased packing efficiency (horn aperture area/total area) of 0.1.

\begin{figure}[h!]
 \centering
	\includegraphics[angle=0,width=0.45\textwidth,]{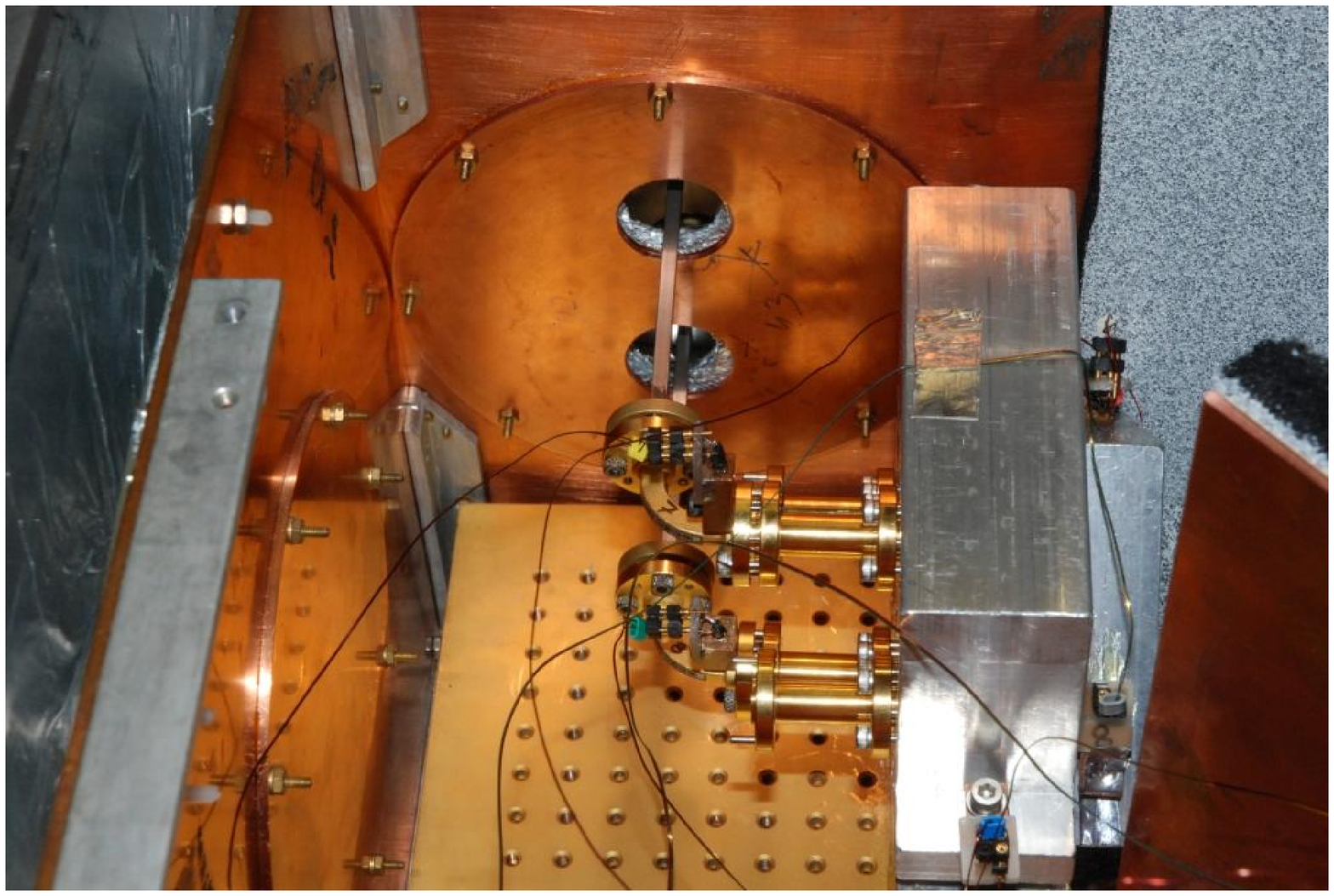}
	\includegraphics[angle=0,width=0.45\textwidth]{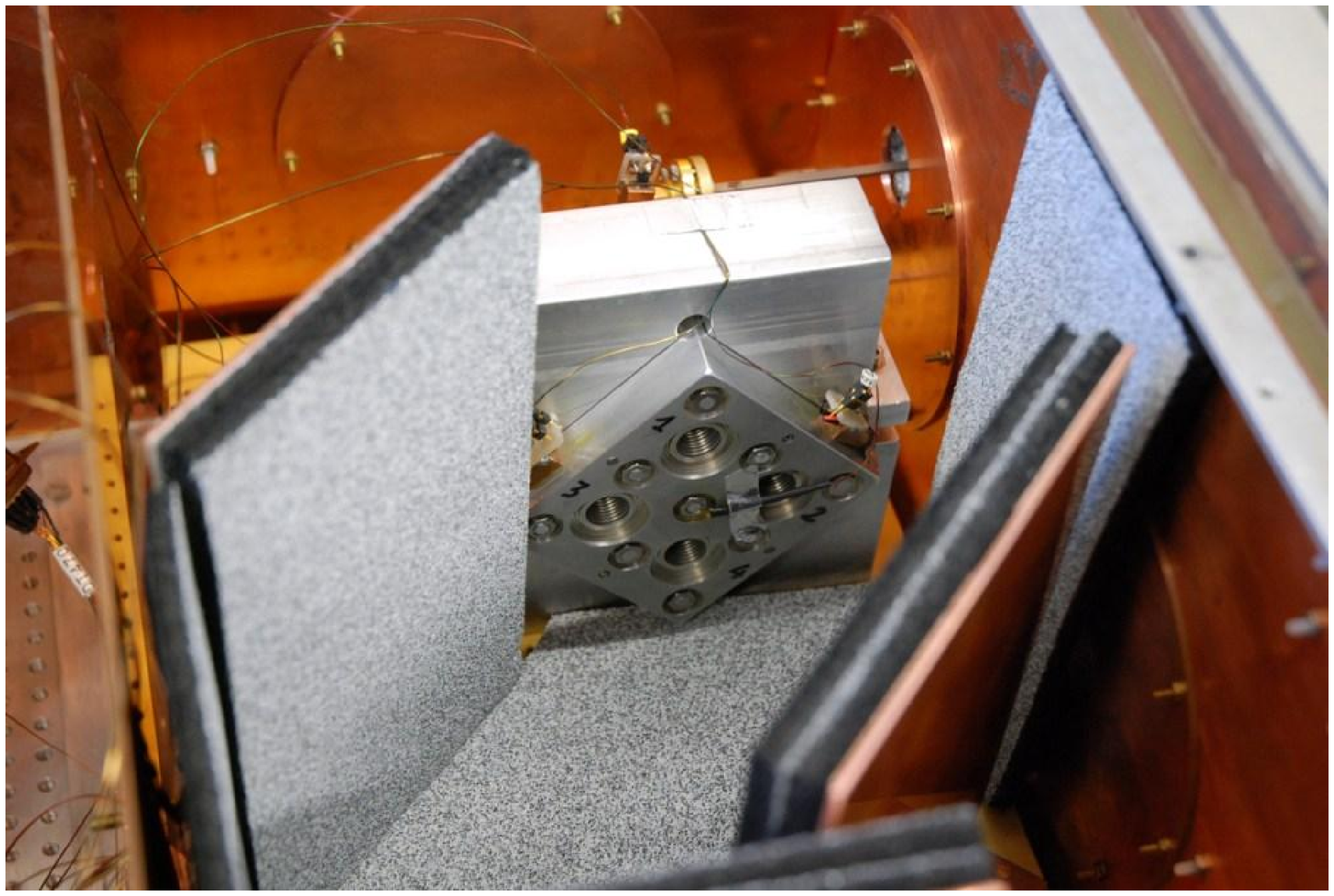}
	 \caption{The cryogenic setup used for return loss measurements. Left: the waveguides chain inside the cryogenic chamber connecting the prototype to the waveguide vacuum window. Right: the horn array with thermometers, facing an ECCOSORB$^\copyright$ AN 72 electromagnetic trap to absorb transmitted radiation.}
	\label{fig:cryo_setup}
\end{figure}

The RL was measured, for one horn, also in cryogenic conditions, after two cooling cycles at about 4 K in a cryogenic chamber. The horn, inside the cryogenic chamber, was connected to the VNA port by means of a waveguides chain (Figure \ref{fig:cryo_setup}) composed by (from the horn to the VNA):  i) a tapered rectangular-to-circular transition, ii) a waveguide bend, iii) a straight steel waveguide (acting as a thermal break), iv) a waveguide vacuum window and v) an external straight waveguide. 

\begin{figure}[!ht]
\centering
	\includegraphics[angle=0,width=0.85\textwidth]{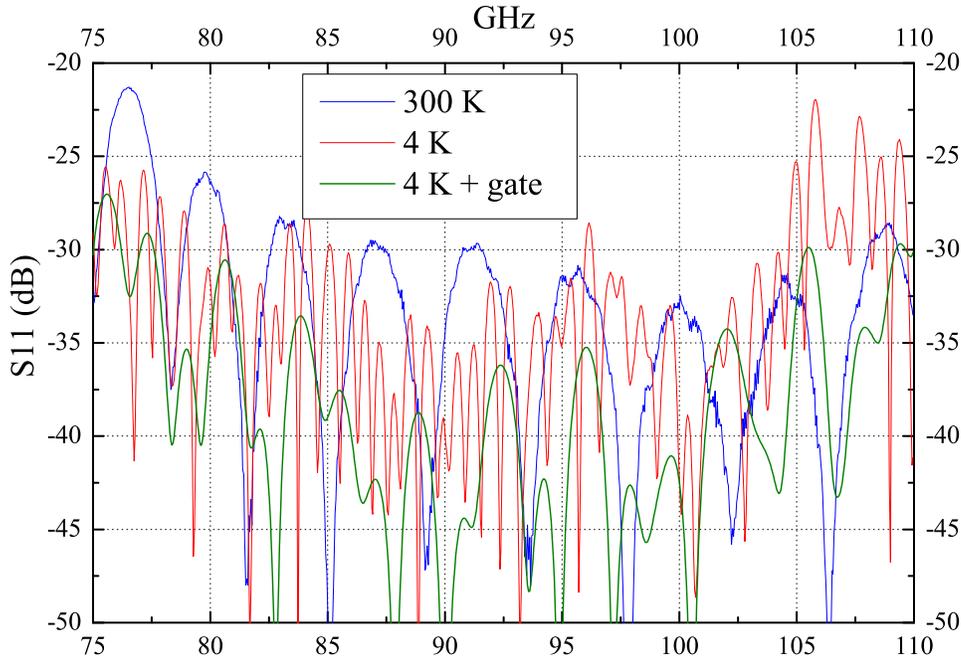}
	 \caption{Comparison of the RL of one horn at 300 K (blue curve) and 4 K (red and green curves). The blue and red curves include also the return loss of the (same) waveguides chain, that connects the horn inside the cryogenic chamber to the external VNA. The waveguides chain RL is responsible for the oscillatory signature. In the green curve, the waveguide chain was filtered with a gate in the VNA time domain. These measurements do not indicate a major change in the return loss that would otherwise occur if there were thermally induced changes in the array.}
	\label{fig:4KRL}
\end{figure}

The VNA was calibrated with a standard one-port calibration at the interface between the last waveguide and the vacuum window. The prototype was positioned inside the cryogenic chamber and connected to the waveguides chain. The RL was first measured at 300 K and then at 4 K. For this reason the measurements, shown in blue and red in Figure \ref{fig:4KRL}, include the return loss of the chain connecting the horn to the VNA. The waveguides chain RL is responsible for the oscillatory signature. By applying a ``gate'' (i.e. a filter in the time domain) with the VNA, the effect of the waveguide chain is removed from the measurements (green line in Fig.~\ref{fig:4KRL}).

It is worth noting that the filtered measurements are significantly better than those performed in standard conditions, because the gate removes the effect of the signal reflected by the waveguide chain. Considering the ``gated'' measurements, the return loss is still below -30 dB on a 30\% bandwidth, confirming the initial room temperature performance. In any case, both red and green curves, do not indicate any mismatch due to possible thermally induced plate displacements. This is an important result confirming the platelet technique reliability.

In Figure \ref{fig:4KRL} one can notice a return loss increase in the 105-110\,GHz region between 300\,K and 4\,K. This effect could be due to thermally induced changes in the waveguides chain, in the absorbing trap and/or in the platelet array itself. It is reasonable to assume that the main contribution comes from the waveguides chain, that is composed by two golden copper waveguides and a straight steel waveguide, acting as a thermal break. The steel waveguide, in particular, has a very different thermal expansion coefficient compared to the other waveguide sections and to the platelet array. Moreover the steel waveguide is the longest one and is subject to the largest thermal gradient which causes the largest differential thermal contraction. These effects are likely to be the main cause of the oscillatory signature. Also thermal effects in the absorbing trap and in the platelet array cannot be excluded but they are likely to be much smaller than those in the waveguides chain.

\subsection{Insertion Loss}
\begin{figure}[!ht]
\centering
	\includegraphics[angle=0,width=0.75\textwidth]{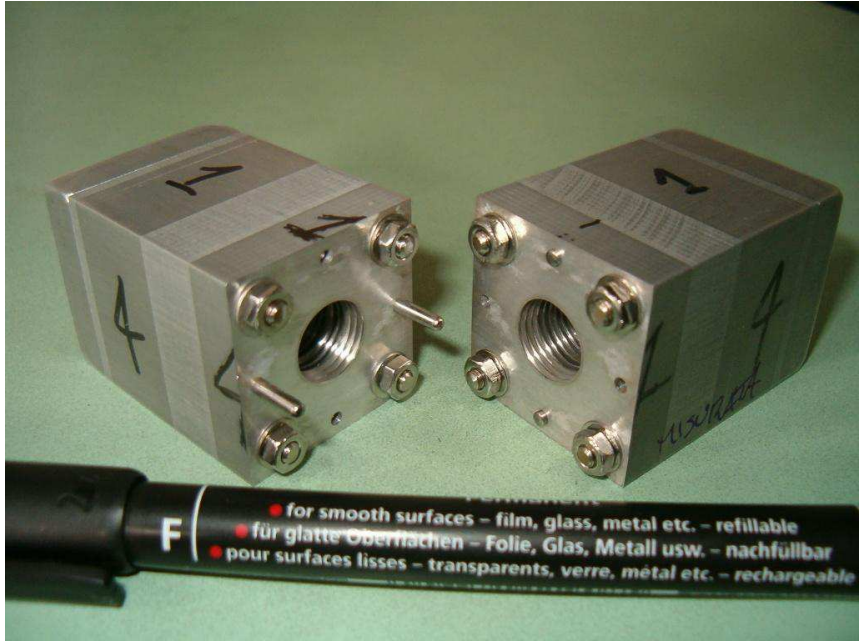}
	 \caption{The two halves, individually re-assembled, of the Twin Horn device, constituted by two identical platelet horns with the same WCAM01 design, used to measure the insertion loss. In the IL measurements the two horns were assembled together with four single rods of double length, with nuts present only on the external sides.}
	\label{fig:TwinHorn}
\end{figure}

The Insertion Loss measurements were performed in transmission, with a TRL calibration, using an Agilent 8510c VNA at the Physics Department of the University of Milano-Bicocca. The signal was injected inside the horn throat and measured at its exit. The difference between the output and injected signals yields the lost signal absorbed by the material because of its non-perfect electrical conductivity and backscattered at the input (the return loss, that is negligible in this case). In order to collect the output signal, we used a second twin horn, coupled by the aperture to the first. Because two horns were used, a factor 2 must be taken into account in order to obtain the single horn IL. Furthermore we simplified the mechanical interface by realizing a dedicated pair of ``twin horns'' as a single two-port device constituted by two faced identical horns with the same design of the array horns and manufactured according to the same technique (Figure \ref{fig:TwinHorn}). 

The insertion loss measurements were obtained by connecting the two horns to the two Agilent 8510c VNA ports and measuring the $S_{2,1}$ (or $S_{1,2}$) parameter with the same rectangular-to-circular transitions used for the RL measurements. The VNA was calibrated with a TRL calibration at the circular interface between the horns and the transitions. 

\begin{figure}[!ht]
\centering
	\includegraphics[angle=0,width=0.85\textwidth]{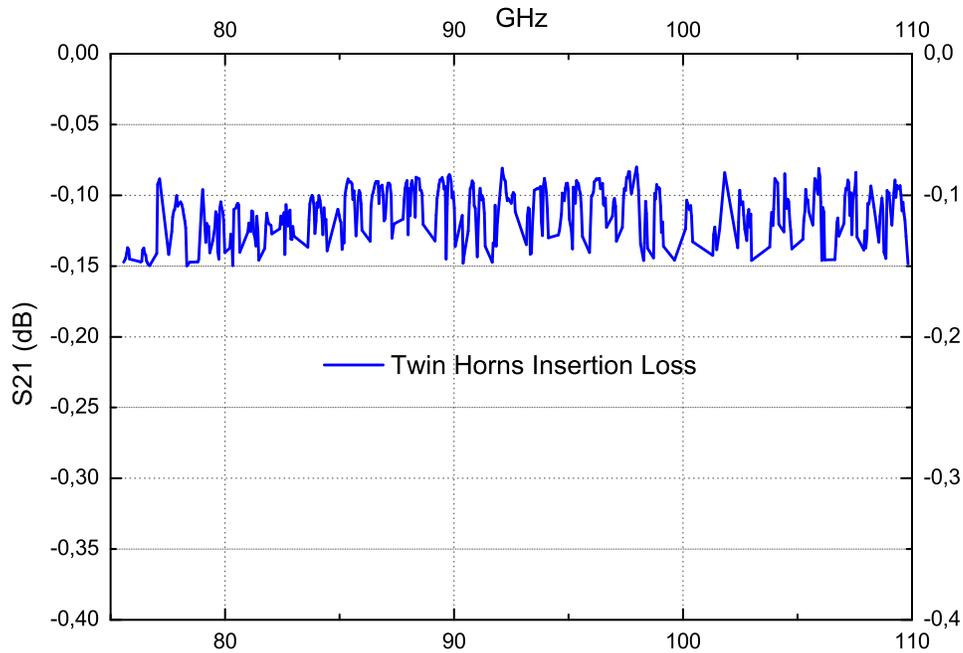}
	 \caption{Insertion Loss of the Twin Horns measured in transmission (S21). Data were resampled in order to eliminate negative resonance spikes. The insertion loss of a single horn is the half of that shown in the figure.}
	\label{fig:IL_P2_L1}
\end{figure}

The IL measurement is shown in Figure \ref{fig:IL_P2_L1}, yielding an average value over the band of about -0.06~dB per single horn. It is worth underlining that this figure represent an upper limit, because the wave fronts of the two horns do not match perfectly, being spherical, an not planar, at their mouth. This slight mismatch causes resonances between the horns that were removed by downsampling the measurements (see Fig.~\ref{fig:IL_P2_L1}) but yield, nevertheless, a slightly larger average insertion loss.

%%%%%%%%%%%%%%%%%%%%%%%%%%%%%%%%%%%% CONCLUSIONS %%%%%%%%%%%%%%%%%%%%%%%%%%%%%%%%%%%%

\section{Conclusions}
In this paper we have discussed the manufacturing and testing of a non-bonded W-band 2$\times$2 array prototype of corrugated feed-horns realised by ``platelet'' technique. Feed-horns have been realized by layers, allowing reduced times and costs with respect to standard techniques like electroformation, typically adopted for frequencies $\geq$ 100~GHz.

Beam patterns and RL measurements are in good agreement with the corresponding simulations and show high repeatability for all the feed-horns, also in cryogenic conditions. In particular the RL is less than -30\,dB over a 30\% bandwidth with a maximum matching of -42\,dB at 100\,GHz for all four antennas. Also beam pattern measurements indicate very good repeatability and a low level of side lobes and cross-polar polarisation (both lower than -25\,dB). Moreover an estimation of the IL indicates that it is not worse than -0.06 dB over the W band.

These results are very promising for the time and cost-effective mass production of large array of high-performances corrugated feed-horns, as showed by \cite{Wollack}. The non-bonded solution, chosen here and in \cite{Kangas} to assemble the module, allow simple manufacturing with the possibility to disassemble the module if needed and maintain electromagnetic performance at cryogenic conditions. Further developments of this assembling solution are needed (and in progress) for increasing the packing efficiency (horn aperture area/total area) to 0.3 and reducing the mass. These developments will enable the realization of large, lightweight and high density focal planes, as required by the next generation experiments for the precise measurements of the CMB polarisation.

%%%%%%%% Acknowledgments %%%%%%%%%%
%%%%%%%%%%%%%%%%%%%%%%%%%%%%%%%%%%%
\section{Acknowledgments}
The work described in this paper was carried out in the context of PRIN-MIUR 2006 and ASI contract I/016/07/0 ``COFIS''.

%%%%%%%% Bibliografia %%%%%%%%%%%%%
%%%%%%%%%%%%%%%%%%%%%%%%%%%%%%%%%%%

%%%%%%%%%%%%%%%%%%%%%%%%%%%%%%%
\end{document}